\documentclass[twoside,12pt]{article}  %%% ���� LaTeX ��ʽ�ļ�������ָ�����ȱ��
\usepackage{CJK}   %%%%% ʹ�����ġ����ġ�����ʱ���ô˺���
\usepackage{indentfirst}  %% �ڱ�������һ�ж�����
\usepackage{bm}    %%%%%  ��ӡ��ѧ���ź�б��ʱ���ô˺���
\usepackage{graphicx}  %%%% ���в��� eps ͼ��ʱ���ô˺�������ͼ���� 1
\begin{document}    %% �ı��ļ���ʼ�����Ǳ�����ָ��

\begin{CJK*}{GBK}{song}  %% ��ʼ�������Ļ���

%=================== Text begin here =============================================

\begin{center}
\LARGE\bf Quantum coherence and non-Markovianity of atom in dissipative cavity under weak measurement$^{*}$   %% ������Ŀ
\end{center}

\footnotetext{\hspace*{-.45cm}\footnotesize $^*$Project supported by the Scientific Research Project of Hunan Provincial Education Department, China (Grant No 16C0949), Hunan Provincial Innovation Foundation for Postgraduate (CX2017B177), the National Science Foundation
of China (Grant No 11374096) and the Doctoral Science Foundation of Hunan Normal University, China}

\footnotetext{\hspace*{-.45cm}\footnotesize $^\dag$Corresponding author. E-mail:zhmzc1997@126.com, tel:13807314064}

\begin{center}
\rm Liu Yu ,  \  Zou Hong-Mei$^{\dagger}$ \ and \ Fang Mao-Fa
\end{center}

\begin{center}
\begin{footnotesize} \sl
Key Laboratory of Low-dimensional Quantum Structures and Quantum Control of Ministry of Education, College of Physics and Information Science, Hunan Normal University, Changsha, 410081, China\  \\   %%%% ��ַ a)
%%% ������ַ������������
\end{footnotesize}
\end{center}

\begin{center}
\footnotesize (Received X XX XXXX; revised manuscript received X XX XXXX)
          %% (Received �� �� ��; revised manuscript received �� �� ��)
\end{center}

\vspace*{2mm}

\begin{center}
\begin{minipage}{15.5cm}
\parindent 20pt\footnotesize
%%%% ����ժҪ
Quantum coherence and non-Markovianity of an atom in dissipative cavity under weak measurement are investigated in this work. We find that, the quantum coherence obviously depends on the atomic initial state, the strength of the weak measurement and its reversal, the atom-cavity coupling constant and the non-Markovian effecct. The more obvious the weak measurement effect is, the better the protection of coherence is. The quantum coherence is preserved more efficiently for lager the atom-cavity coupling. The stronger the non-Markovian effect, the more slowly the coherence reduces. This is, the quantum coherence can be effectively protected by means of controlling these physical parameters.
\end{minipage}
\end{center}

\begin{center}
\begin{minipage}{15.5cm}
\begin{minipage}[t]{2.3cm}{\bf Keywords:}\end{minipage}
\begin{minipage}[t]{13.1cm}
%%%%% �ؼ���
quantum coherence, non-Markovianty, dissipative cavity
\end{minipage}\par\vglue8pt
{\bf PACS: }03.65.Yz, 34.80.Qb, 42.25.Kb, 03.65.Ta
%%% PACS ������
%% ��ѯ��ַ��http://www.aip.org/pacs
\end{minipage}
\end{center}

\section{Introduction}  %%% �ڱ��� 1
Quantum coherence marks the departure of quantum mechanics from classical physics. It is increasingly recognized as a vital physical resource in both quantum theory and quantum information. There are several measure methods about quantum coherence, such as the relative entropy of coherence $C_{rel.ent}(\hat{\rho})=S(\hat{\rho}_{diag})-S(\hat{\rho})$, where $S$ is the von Neumann entropy and $\hat{\rho}_{diag}$ denotes the state obtained from $\hat{\rho}$ by deleting all off-diagonal elements, and the intuitive $l_{1}$ norm of coherence $C_{l_{1}}(\hat{\rho})=\sum\limits_{i,j \\ (i\neq j) } |\varrho_{i,j}|$ \cite{T.Baumgratz}. A growing number of applications can be certified to rely on various incarnations of quantum coherence as a primary ingredient. This has been the main motivation for recent researches to quantify and develop a number of measures of quantum coherence. Quantum coherence has been linked with other quantum resources, such as non-locality\cite{K.Bu,D.Mondal,L.Qiu}, non-markovianity\cite{T.Chanda,Z.Huang,Ali Mortezapour}, entanglement\cite{Z.Xi,S.Chin,A.Streltsov}, quantum discord\cite{Y.Yao,Y.Guo}.

Recently, quantum coherence has been established as an important notion. The quantum coherence has been gaining a great deal of attention as a signature of quantum nature and a resource for quantum technologies. For example, Chiranjib \emph{et al.}\cite{Chiranjib} explored quantum coherence in superposition with non-distinguishable pointers, it is plausible that the results will be useful in quantum multiple-slit experiments and interferometric set-ups with feedback loops. The authors in Ref.\cite{Ali Mortezapour} studied non-Markovianity and coherence of a moving qubit inside a leaky cavity. Recently, the applications of weak measurement\cite{Y.Aharonov,M.Koashi} have been proposed as a practical method to protect the quantum coherence\cite{Haozhen}. The weak measurement and its reversal can be utilized to improve the fidelity of system. In this paper, we study the quantum coherence of an atom in dissipative cavity under weak measurement. We find that the quantum coherence can be effectively protected by adjusting the strength of the weak measurement and its reversal, the atom-cavity coupling and the spectral width of the reservoir. Then, we also analyze the effect of the qubit on the degree of non-Markocianity.

The paper is organized as follows. In Section 2, we give a Hamiltonian of an atom in the dissipative cavity and determine a state evolution of the atom referring to this Hamiltonian under weak measurement. In Section 3, we discuss about the coherence of the atom and the influence of the non-Markovianity. Finally, a brief summary is presented in Section 4.

\section{Physical Model}
\subsection{Atom in dissipative cavity}  %%% �ڱ��� 2
Consider an atom interacting with a cavity, where the cavity is coupled to a bosonic environment\cite{ZouHM2}. The Hamiltonian reads

\begin{eqnarray}\label{EB01}
H&=&\frac{1}{2}\omega_{0}\sigma_{z}+\omega_{0}a^{\dag}a+\Omega(a\sigma_{+}+a^{\dag}\sigma_{-})+\sum_{k}\omega_{k}c_{k}^{\dag}c_{k}+(a^{\dag}+a)\sum_{k}g_{k}(c_{k}^{\dag}+c_{k}),
\end{eqnarray}

where $a^{\dag}$ and $a$ are the creation and annihilation operators of the cavity, $\sigma_{\pm}$ and $\omega_{0}$ are the transition operators and Bohr frequency of the atom\cite{Jaynes}, $\Omega$ is the coupling constant between the atom and its cavity, and $c_{k}^{\dag}$ and $c_{k}$ are the creation and annihilation operators of the reservoir. The coupling constant between the cavity and its environment is denoted by $g_{k}$.

Assuming one initial excitation and the reservoir at zero temperature, the non-Markovian master equation for the density operator $R(t)$ in the dressed-state basis $\{|E_{1+}\rangle, |E_{1-}\rangle, |E_{0}\rangle\}$ is

\begin{eqnarray}\label{EB02}
\dot{R}(t)&=&-i[H_{JC},R(t)]\nonumber\\
&+&\gamma(\omega_{0}+\Omega,t)(\frac{1}{2}|E_{0}\rangle\langle E_{1+}|R(t)|E_{1+}\rangle\langle E_{0}|-\frac{1}{4}\{|E_{1+}\rangle\langle E_{1+}|,R(t)\})\nonumber\\
&+&\gamma(\omega_{0}-\Omega,t)(\frac{1}{2}|E_{0}\rangle\langle E_{1-}|R(t)|E_{1-}\rangle\langle E_{0}|-\frac{1}{4}\{|E_{1-}\rangle\langle E_{1-}|,R(t)\}),
\end{eqnarray}
where $|E_{1\pm}\rangle=(|1g\rangle\pm|0e\rangle)/\sqrt{2}$ are the eigenstates of $H_{JC}$ with one total excitation, with energy $\omega_{0}/2\pm\Omega$, and $|E_{0}\rangle=|0g\rangle$ is
the ground state, with energy $-\omega_{0}/2$. The timedependent decay rates for $|E_{1-}\rangle$ and $|E_{1+}\rangle$ are $\gamma(\omega_{0}-\Omega,t)$ and $\gamma(\omega_{0}+\Omega,t)$,
respectively.

If the reservoir at zero temperature is modeled with a Lorentzian  spectral density\cite{ZouHM,zhm,zhm2}

\begin{equation}\label{EB03}
J(\omega)=\frac{1}{2\pi}\frac{\lambda_{0}\lambda^{2}}{(\omega_{1}-\omega)^{2}+\lambda^{2}},
\end{equation}
where $\lambda$ defines the spectral width of the coupling, which is connected to the reservoir correlation time and $\lambda_{0}$ is the system-environment coupling strength. For a weak regime we mean the case $\lambda>2\lambda_{0}$, in this regime the behavior of dynamical evolution of the system is essentially a Markovian exponential decay controlled
 by $\lambda_{0}$. In the strong coupling regime, for $\lambda<2\lambda_{0}$, non-Markovian effects become relevant\cite{Bellomo1,Bellomo2,Breuer1}. We consider the spectrum is peaked on the frequency of the state $|E_{1-}\rangle$,
 i.e., $\omega_{1}=\omega_{0}-\Omega$, the decay rates for the two dressed states $|E_{1\pm}\rangle$ are respectively expressed
 as\cite{Scala1} $\gamma(\omega_{0}-\Omega,t)=\gamma_{0}(1-e^{-\lambda t})$ and $\gamma(\omega_{0}+\Omega,t)=\frac{\lambda_{0}\lambda^{2}}{4\Omega^{2}+\lambda^{2}}\{1+
 [\frac{2\Omega}{\lambda}
 \sin2\Omega t-\cos2\Omega t]e^{-\lambda t}\}$.

In the dessed-state basis, if $R_{ij}(0)(i,j=1,2,3)$ describes the matrix elements of the initial atom-cavity state, we can acquire the matrix elements at all times from Eq.(2)

\begin{eqnarray}\label{EB4}
      R_{11}(t)&=&A_{11}^{11}R_{11}(0),      R_{12}(t)=A_{12}^{12}R_{12}(0),      R_{13}(t)=A_{13}^{13}R_{13}(0),\nonumber\\
      R_{22}(t)&=&A_{22}^{22}R_{22}(0),      R_{23}(t)=A_{23}^{23}R_{23}(0),\nonumber\\
      R_{33}(t)&=&A_{33}^{11}R_{11}(0)+A_{33}^{22}R_{22}(0)+A_{33}^{33}R_{33}(0),
\end{eqnarray}
here,
\begin{eqnarray}\label{EB5}
      A_{11}^{11}&=&e^{-\frac{1}{2}I_{+}},   A_{12}^{12}=e^{-2i\Omega t}e^{-\frac{1}{4}(I_{+}+I_{-})}, A_{13}^{13}=e^{-i(\omega_{0}+\Omega)t}e^{-\frac{1}{4}I_{+}},\nonumber\\
      A_{22}^{22}&=&e^{-\frac{1}{2}I_{-}},   A_{23}^{23}=e^{-i(\omega_{0}-\Omega)t}e^{-\frac{1}{4}I_{-}},\nonumber\\
      A_{33}^{11}&=&1-A_{11}^{11},   A_{33}^{22}=1-A_{22}^{22},      A_{33}^{33}=1,
\end{eqnarray}
and

\begin{eqnarray}\label{EB6}
     I_{-}&=&\lambda_{0}t+\frac{\lambda_{0}}{\lambda}(e^{-\lambda t}-1),\nonumber\\
     I_{+}&=&\frac{\lambda_{0}\lambda^{2}}{4\Omega^{2}+\lambda^{2}}[t-\frac{4\Omega e^{-\lambda t} \sin(2\Omega t)}{4\Omega^{2}+\lambda^{2}}+\frac{(\lambda^{2}-4\Omega^{2})(e^{-\lambda t}
     \cos(2\Omega t)-1)}{\lambda(4\Omega^{2}+\lambda^{2})}].
\end{eqnarray}

We consider the coherence of single atom in dissipative cavity. The atom desity operator in the standard state is

\begin{eqnarray}\label{EB7}
      \rho&=&\left(
                   \begin{array}{cccc}
                    \rho_{11}(t)&\rho_{12}(t)\\
                    \rho_{21}(t)&\rho_{22}(t)\\
                    \end{array}
              \right),
\end{eqnarray}
where
\begin{eqnarray}\label{EB8}
    \rho_{11}(t)&=&\frac 12 (R_{11}(t)-R_{12}(t)-R_{21}(t)+R_{22}(t)),\nonumber\\
    \rho_{12}(t)&=&\frac 12 (\sqrt 2 R_{13}(t)-\sqrt 2 R_{23}(t)),\nonumber\\
    \rho_{21}(t)&=&\frac 12 (\sqrt 2 R_{31}(t)-\sqrt 2 R_{32}(t)),\nonumber\\
    \rho_{22}(t)&=&\frac 12(R_{11}(t)-R_{12}(t)+R_{21}(t)+R_{22}(t)+2R_{33}(t)).\nonumber\\
\end{eqnarray}

\subsection{weak measurement}  %%% �ڱ��� 3
In the dissipative cavity, the weak measurement and its reversal\cite{Ming} can be applied in two places: one before and the other after the atom interacting with the dissipative cavity. The operator of the weak measurement can be described as

\begin{eqnarray}\label{EB9}
    b_{1}&=&\left(
                  \begin{array}{cccc}
                  1&0\\
                  0&\sqrt{1-p_{1}}\\
                  \end{array}
             \right),
\end{eqnarray}
where $p_{1}$ is the strength of the weak measurement. The bigger the $p_{1}$ value, the stronger the weak measurement, the weak measurement become the strong measurement when $p_{1}=1$.
Let the initial state be

\begin{eqnarray}\label{EB10}
    \varphi_{0}&=&\cos{\frac{\theta}{2}}|e\rangle+\sin{\frac{\theta}{2}}|g\rangle,
\end{eqnarray}
the atomic state after the weak measurement can be described as
\begin{eqnarray}\label{EB11}
    \rho'(0)&=&b_{1}\rho(0)b_{1}^{\dag}=\left(
                  \begin{array}{cccc}
                  \rho'_{11}(0)&\rho'_{12}(0)\\
                  \rho'_{21}(0)&\rho'_{22}(0)\\
                  \end{array}
             \right)\nonumber\\
    &=&\left(
                  \begin{array}{cccc}
                  \sin^{2}(\frac{\theta}{2})&\sqrt{1-p_{1}}\sin(\frac{\theta}{2})\cos(\frac{\theta}{2})\\
                  \sqrt{1-p_{1}}\sin(\frac{\theta}{2})\cos(\frac{\theta}{2})&(1-p_{1})\cos^{2}(\frac{\theta}{2})\\
                  \end{array}
             \right).
\end{eqnarray}
 Then, the atom interacts with the dissipative cavity, so the atom state at time $t$ can be described
 \begin{eqnarray}\label{EB12}
     \rho'(t)&=&\left(
                  \begin{array}{cccc}
                  \rho'_{11}(t)&\rho'_{12}(t)\\
                  \rho'_{21}(t)&\rho'_{22}(t)\\
                  \end{array}
             \right).
 \end{eqnarray}
From Eq.(4), (8) and (11), we can get the matrix elements as
 \begin{eqnarray}\label{EB13}
         \rho'_{11}(t)&=&\frac{1}{4}(A^{11}_{11}+A^{12}_{12}+A_{21}^{21}+A_{22}^{22})\rho'_{11}(0),\nonumber\\
         \rho'_{12}(t)&=&\frac 1 2 (A^{13}_{13}+A^{23}_{23})\rho'_{12}(0),\nonumber\\
         \rho'_{21}(t)&=&\frac 1 2 (A^{31}_{31}+A^{32}_{32})\rho'_{21}(0),\nonumber\\
         \rho'_{22}(t)&=&\frac 1 2 ((\frac 1 2 A_{11}^{11}-\frac 1 2 A_{12}^{12}+\frac 1 2 A_{22}^{22}-\frac 1 2 A_{21}^{21})\rho_{11}(0)+2\rho'_{22}(0)).\nonumber\\
 \end{eqnarray}

The operator of the reversal is given by
 \begin{eqnarray}\label{EB14}
    b_{2}&=&\left(
                  \begin{array}{cccc}
                  \sqrt{1-p_{2}}&0\\
                  0&1\\
                  \end{array}
             \right),
\end{eqnarray}
where $p_{2}$ is the strength of the reversal.

The atom state can be described as
\begin{eqnarray}\label{EB15}
    \rho^{w}(t)&=&b_{2}\rho'(t)b_{2}^{\dag}=\left(
                  \begin{array}{cccc}
                  \rho^{w}_{11}(t)&\rho^{w}_{12}(t)\\
                  \rho^{w}_{21}(t)&\rho^{w}_{22}(t)\\
                  \end{array}
                  \right)
                  =
    \left(
                  \begin{array}{cccc}
                 (1-p_{2})\rho'_{11}(t)&\frac 1 2\sqrt{1-p_{2}}\rho'_{21}(t)\\
                 \frac 1 2 \sqrt{1-p_{2}\rho'_{21}(t)}&\rho'_{22}(t)\\
                  \end{array}
             \right).
\end{eqnarray}

 In the following, we calculate the quantum coherence of the atom by using the intuitive $\ell$ norm of coherence, it is
 \begin{eqnarray}\label{EB16}
 C_{\rho}&=\sum\limits_{i,j(i\neq j)}|\rho_{i,j}^{w}|=\rho^{w}_{12}(t)+\rho^{w}_{21}(t),
 \end{eqnarray}
 where $\rho^{w}_{12}(t)$ and $\rho^{w}_{21}(t)$ are the off-diagonal elements of the system density matrix, respectively.

\section{Discussions and Results}
\subsection{Quantum coherence }  %%% �ڱ��� 4
To investigate the quantum coherence evolution of the atom in the dissipative cavity, we numerically calculate the quantum coherence and analyze the behavior of the coherence in different condition.
Figure 1 shows the behavior of $C_{\rho}$ as a function of the initial state $|\varphi_{0}\rangle$ . In Figure 1 we choose as $\Omega=\lambda_{0}$, $\lambda=5\lambda_{0}$,
$\lambda_{0}t=10$, $p_{1}=0.5$ and $p_{2}=0.5$. The result clearly shows that the coherence changes cyclically. In addition the coherence get the maximum value at $\theta=\frac{\pi}{2}$ or $\frac{3\pi}{2}$, i.e., when the initial state is $|\varphi_{0}\rangle $=$\frac{1}{\sqrt{2}}(|e\rangle+|g\rangle)$, there is the best coherence.

We depict the coherence as a function of parameters $p_{1}$ and $p_{2}$ in Figure 2 . One can clearly see that the behavior of the coherence $C_{\rho}$ depends on the parameters $p_{1}$ and $p_{2}$. The numerical results indicate that the coherence decreases by increasing the strength of the weak measurement or its reversal. This means that the better coherence is maintained when weak measurement is more obvious. And in the optimal initial state, $p_{1}=0$ and $p_{2}=0$, the atom has the best coherence.

\begin{figure}[htbp]
\begin{minipage}[t]{0.4\linewidth}
\centering
\includegraphics[width=7.5cm,height=4.5cm]{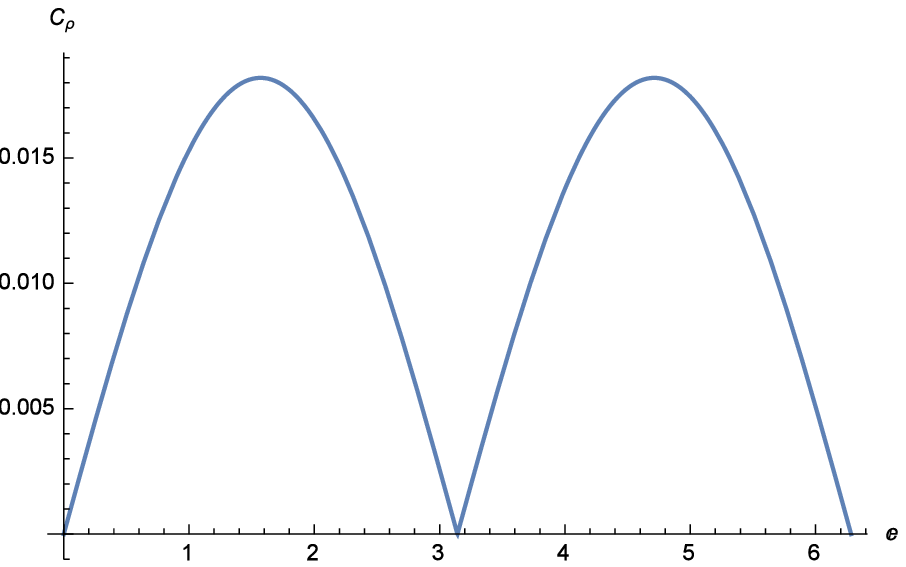}
\caption{we chosen as $\Omega=\lambda_{0}$, $\lambda=5 \lambda_{0} $, $\lambda_{0}t=10$ and $p_{1}=p_{2}=0.5$.\label{fig:side:a}}
\end{minipage}%
\hfill
\begin{minipage}[t]{0.5\linewidth}
\centering
\includegraphics[width=7.5cm,height=4.5cm]{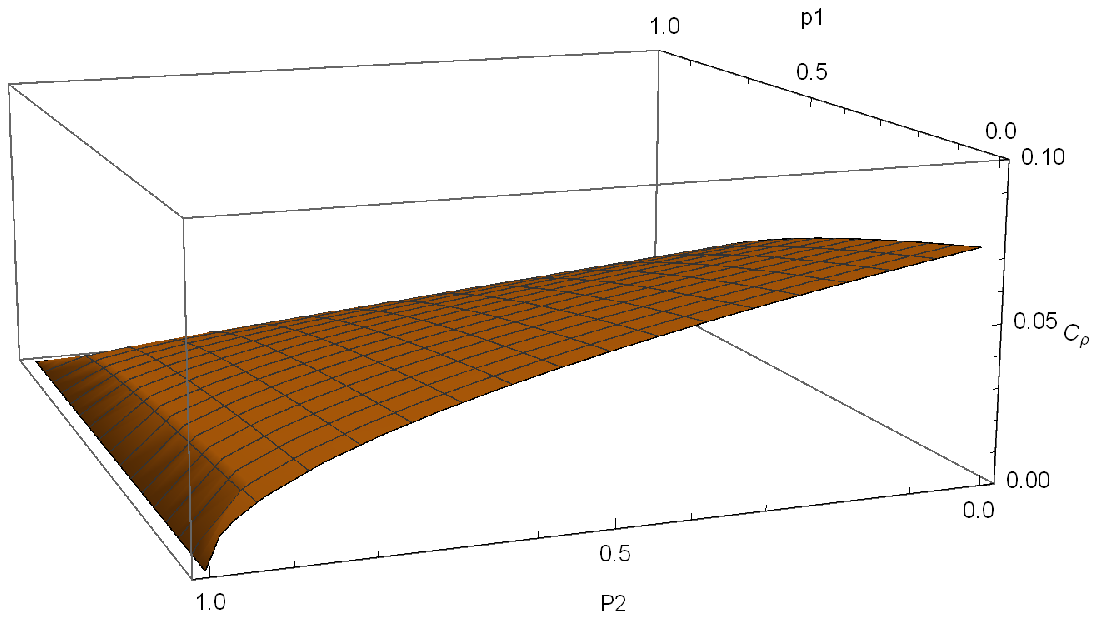}
\caption{We chosen as $\Omega=\lambda_{0}$, $\lambda=5\lambda_{0}$. The initial state chosen
$|\varphi_{0}\rangle $=$\frac{1}{\sqrt{2}}(|e\rangle+|g\rangle) $\label{fig:side:b}}
\end{minipage}
\end{figure}

Figure 3 and Figure 4 are the image of quantum coherence as the functions of $\Omega/\lambda_{0}$. If $\lambda=3\lambda_{0}$, the reservoir is in the Markovain regime. For the cavity is dissipative, the coherence will decreases with time. In Figure 3, we find that the coherence almost reduces to zero when the coupling constant between the atom and its cavity is small. The decay of coherence is getting slower when $\Omega$ increases. In particular, if the coupling constant is large enough, the coherence tends to a finite value after a long time. Figure 4 is a more detailed description of the quantum coherence as the functions of $\Omega$. It can clearly display the influence of coupling constant $\Omega$ on the decay behavior of the coherence. Figure 4 indicates that, the coherence quickly reduces to zero when $\Omega=\lambda_{0}$, however, the coherence slowly decays to zero when $\Omega=10\lambda_{0}$. But, the coherence is finally approach to 0.36 when $\Omega=40\lambda_{0}$. The physical explanation is that, the interaction between the atom and its cavity is large enough, so that the impact of the reservoir on the atom is relatively weak. The results show the quantum information will be imprisoned in the atom-cavity and the quantum coherence have been effectively  protected.

\begin{figure}[htbp]
\begin{minipage}[t]{0.4\linewidth}
\centering
\includegraphics[width=7.5cm,height=4.5cm]{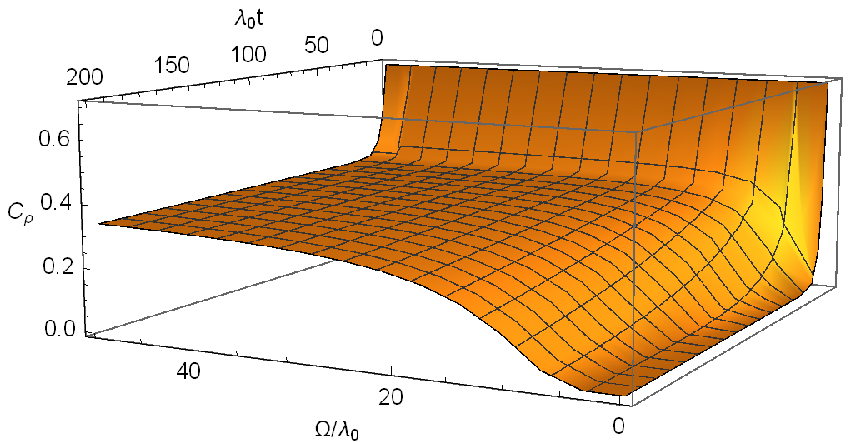}
\caption{The coherence as a function of the coupling constant $\Omega$ and time $\lambda_{0}t$. We chosen as $p_{1}=0$, $p_{2}=0$, $\lambda=3\lambda_{0}$. The initial state chosen
$|\varphi_{0}\rangle $=$\frac{1}{\sqrt{2}}(|e\rangle+|g\rangle) $.\label{fig:side:c}}
\end{minipage}%
\hfill
\begin{minipage}[t]{0.5\linewidth}
\centering
\includegraphics[width=7.5cm,height=4.5cm]{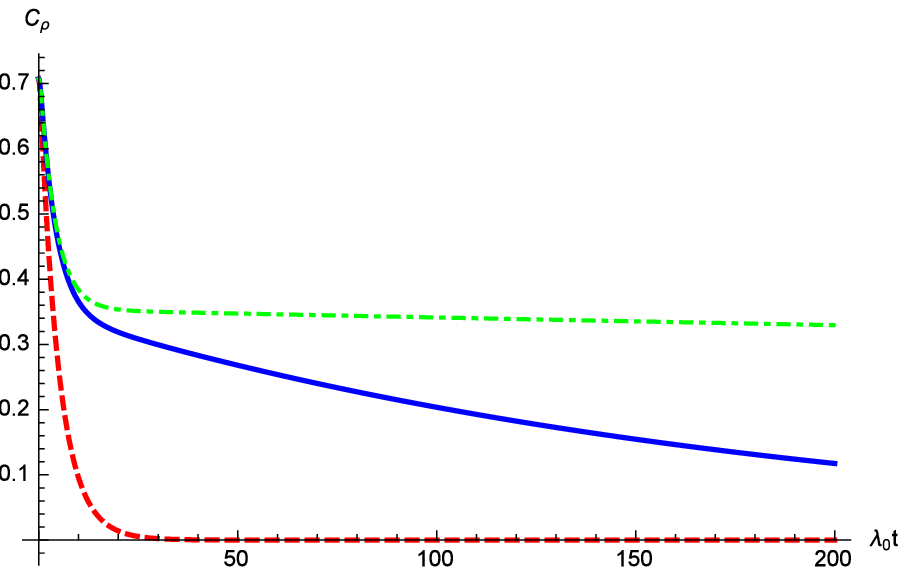}
\caption{(color online). The dotted red line represents, $\Omega=\lambda_{0}$; the solid blue line, $\Omega=10\lambda_{0}$; the dot-dashed green line, $\Omega=40\lambda_{0}$. \label{fig:side:d}}
\end{minipage}
\end{figure}

 We depict the coherence $C_{\rho}$ as a function of $\lambda/\lambda_{0}$ and $\lambda_{0}t$ in the case of $\Omega=\lambda_{0}$ in Figure 5 and Figure 6. Figure 5 clearly shows that the coherence $C_{\rho}$ of the atom in the dissipative cavity will quickly decay to zero when $\lambda>2\lambda_{0}$(in the Markovian regime). For $\lambda<2\lambda_{0}$, the non-Markovian effects become relevant. When $\lambda/\lambda_{0}$ approach to 0, the coherence is well maintained and reach a fixed value. For a further understanding of the coherence $C_{\rho}$ under different coupling strength, Figure 6 demonstrates the coherence $C_{\rho}$ as a function of $\lambda_{0}t$ in different $\lambda$. We can see that the coherence is greatly influenced by the reservoir when $\Omega=\lambda_{0}$. That is to say, the coherence in the non-Markovian regime reveals distinct features: although the coherence will decay to zero in a long time when $\lambda=\lambda_{0}$, the decay rate of coherence in this case is obviously smaller than in Markovian regime($\lambda=3\lambda_{0}$). Comparing $\lambda=0.01\lambda_{0}$ and $\lambda=0.1\lambda_{0}$, it is seen that, the smaller the value of $\lambda$ is, the stronger the non-Markovian effect is, the slower the entanglements reduce. The coherence will tend to 0.36 when the non-Markovian effect is strong enough.

\begin{figure}[htbp]
\begin{minipage}[t]{0.4\linewidth}
\centering
\includegraphics[width=7.5cm,height=4.5cm]{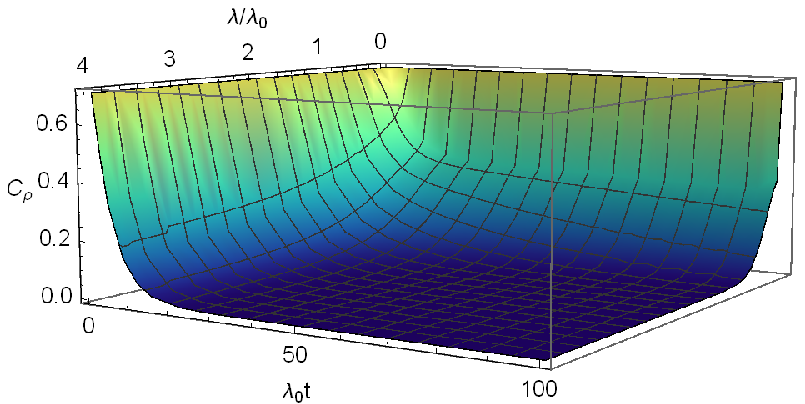}
\caption{The coherence for $\lambda/\lambda_{0}$ given by equation(12).Where the parameter $p_{1}=0$ , $p_{2}=0$ ,$\Omega=\lambda_{0}$. The initial state chosen
$|\varphi_{0}\rangle $=$\frac{1}{\sqrt{2}}(|e\rangle+|g\rangle) $.\label{fig:side:a}}
\end{minipage}
\hfill
\begin{minipage}[t]{0.5\linewidth}
\centering
\includegraphics[width=7.5cm,height=4.5cm]{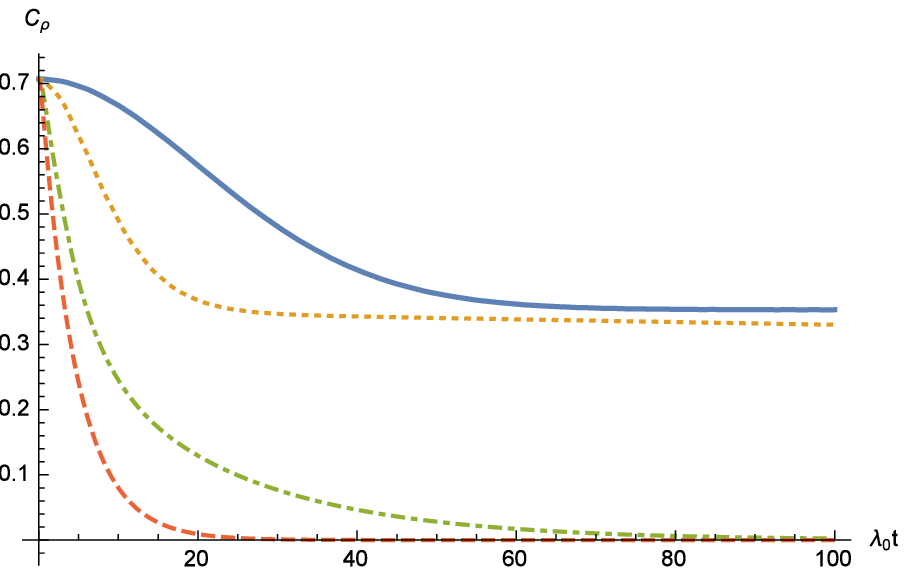}
\caption{(color online). The solid blue line represents, $\lambda=0.01\lambda_{0}$; the dotted yellow line, $\lambda=0.1\lambda_{0}$; the dot-dashed green line, $\lambda=\lambda_{0}$; the dashed redline, $\lambda=3\lambda_{0}$.\label{fig:side:b}}
\end{minipage}
\end{figure}

\subsection{Non-markovianity}
The measure $N(\Phi)$ for non-markovianity of the quantum process $\Phi(\emph{t})$ has been defined by Ref.\cite{H.P.}. Considering a quantum process $\Phi(\emph{t})$, this quantity depends on time $t$ and on the intitial states $\rho_{1,2}(0)$ with corresponding time evolutions $\rho_{1,2}(t)=\Phi(t,0)\rho_{1,2}(0)$. The non-markovianity $N(\Phi)$ is defined as
\begin{eqnarray}\label{EB17}
     N(\Phi)&=&\max_{\rho_{1,2}(0)}\int_{\sigma>0}\emph{dt}\sigma[\emph{t},\rho_{1,2}],
\end{eqnarray}
where $\sigma[\emph{t},\rho_{1,2}(0)]$ is the rate of change of the trace distance $\sigma[\emph{t},\rho_{1,2}(0)]=\frac{\emph{d}}{dt}\emph{D}[\rho_{1}(\emph{t}),\rho_{2}(\emph{t})]$. The trace distance $D$ describing the distinguish ability between the two states is defined as \cite{M.A.}
\begin{eqnarray}\label{EB18}
           D(\rho_{1},\rho_{2})&=&\frac{1}{2}\|\rho_{1}-\rho_{2}\|,
\end{eqnarray}
where $\|M\|=\sqrt{M^{\dag}M}$ and satisfying $0 \leq D \leq 1$. $\sigma[\emph{t},\rho_{1,2}(0)]\leq 0$ corresponds to all dynamical semigroups and all time-dependent Markovian processes; a process is non-Markovian if there exists a pair of initial state at a certain time $\emph{t}$ such that $\sigma[\emph{t},\rho_{1,2}(0)]>0$. We take the maximum over all initial states $\rho_{1,2}(0)$ to calculate the degree of non-Markovianity. By drawing a sufficiently large sample of random pairs of initial states, the optimal state pair is attained for the initial states are $\rho_{1}(0)=|g\rangle\langle g|$ and $\rho_{2}(0)=|e\rangle\langle e|$.

 \begin{center}
\includegraphics[width=7.5cm,height=4.5cm]{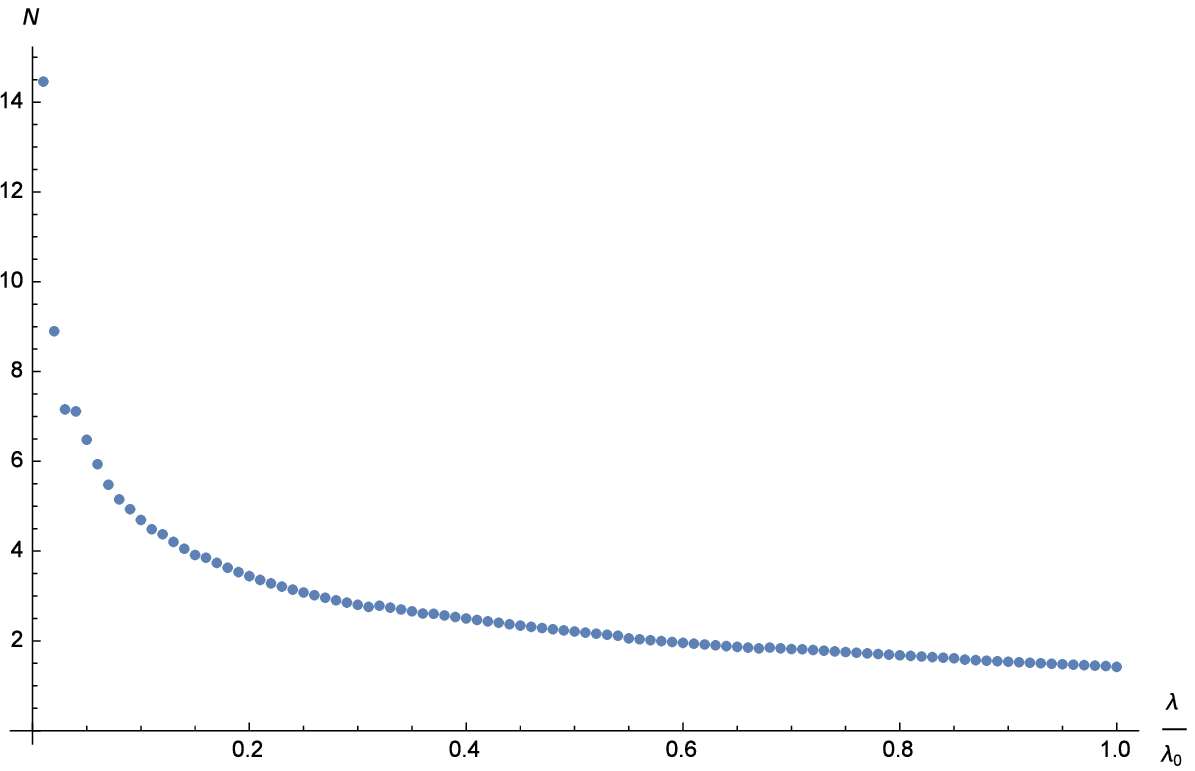}
\parbox{15cm}{\small{\bf Figure7}
(color online). The non-Markovianity $N(\Phi)$ for this model as a function of the $\lambda/\lambda_{0}$. The optimal state pair is attained for the initial states are $\rho_{1}(0)=|g\rangle\langle g|$ and $\rho_{2}(0)=|e\rangle\langle e|$. We choose $p_{1}=0$, $p_{2}=0$ and $\Omega=\lambda_{0}$}
\end{center}

Figure 7 shows the behavior of degree of non-Markovianity as a function of the parameter $\lambda/\lambda_{0}$. From Figure7, we know that $N(\Phi)$ is obviously dependent on  $\lambda/\lambda_{0}$. The bigger the value of $\lambda/\lambda_{0}$ is, the larger the degree of non-Markovianty. Especially, when $\lambda/\lambda_{0}$ increases from 0.01 to 0.1, the damp of $N(\Phi)$ is vary fast but $N(\Phi)$ is still great and is bigger than 5. We again observe the solid blue line($\lambda/\lambda_{0}=0.01$) and the dotted yellow line($\lambda/\lambda_{0}=0.1$) in Figure 6 and we find that their coherence slowly decays to their stable value. In other words, the non-Markovianity could effectively control the coherence decay. With $\lambda/\lambda_{0}$ increasing, $N(\Phi)$ is changed smaller, the coherence will disappear after a long time, shown in Figure 5. When the degree of non-Markovianity is not equal zero there is a flow of information from the environment back to the open system, this is the key feature of non-Markovian dynamics. Therefore, the information is more from reservoir back to the atom-cavity system and the protection of coherence is best when $N(\Phi)$ is lager.

\section{Conclusion}
In this paper we mainly study quantum coherence and non-markovianity of an atom in dissipative cavity under weak measurement. The results show that the quantum coherence obviously depends on the atomic initial state, the strength of the weak measurement and its reversal, the atom-cavity coupling constant and the spectral width of the reservoir. The atom has the best coherence when the atomic initial state is $|\varphi_{0}\rangle $=$\frac{1}{\sqrt{2}}(|e\rangle+|g\rangle $. The weak measurement and its reversal can protect coherence. When $p_{1}\rightarrow0$ and $p_{2}\rightarrow0$, the weak measurement effect is more obvious, therefore the better the protection of coherence.
As the coupling constant $\Omega$ is bigger enough, the coherence will not be affected by the reservoir, i.e., does not decay to 0 or even reaches a fixed value. The smaller the value of $\lambda$ is, the stronger the non-Markovian effect, the more slowly the coherence reduce. The coherence will tend to 0.36 after a long time when the non-Markovian effect is strong enough. On the other hand, the result shows that the degree of non-Markovianity can act as a protector of qubit coherence. The greater degree of non-Markovianity when $\lambda/\lambda_{0}\rightarrow0$. By this time, the information is lager about the reservoir back flow, so the better protect of the quantum coherence of the atom.

\vspace*{2mm}

%%%% �ο������Ű���ʽ��

\end{CJK*}  %% �������ġ����ġ�����ʹ�û���

\begin{thebibliography}{99}
\itemsep=-4pt plus.2pt minus.2pt  %% �����ο�����������֮���ļ���
\small
%\bibitem{1}  %% �����ο�����1����
%\bibitem{2}  %% �����ο�����2����
%\bibitem{3} %% �����ο�����3���ݣ���������������
%%%%����1-10
\bibitem{T.Baumgratz}Baumgratz T, Cramer M and Plenio M B 2014 {\emph{Phys. Rev. Lett } \textbf{113} 140401}
\bibitem{K.Bu}Bu K, Kumar A and Wu J{2016 arXiv:1603.06322 [quant-ph]}
\bibitem{D.Mondal}Mondal D, Pramanik T,and Pati A K {2017 \emph{Phys. Rev. A } \textbf{95} 010301}
\bibitem{L.Qiu}Qiu L, Liu Z and Wu J {2016 arXiv:1610.07237[quan-ph]}
\bibitem{T.Chanda}Chanda T and Bhattacharya S {2016 \emph{Annals of Physics} \textbf{366} 1}
\bibitem{Z.Huang}Huang Z and Situ H {2017 \emph{International Journal of Theoretical Physics}\textbf{56} 503}
\bibitem{Ali Mortezapour}Mortezapour A, Borji M A and Franco R L {2017 arXiv:1705.00887v1 [quant-ph]}
\bibitem{Z.Xi}Xi Z, Li Y and Fan H {2015 \emph{Sci. Rep.}\textbf{5} 10922}
\bibitem{S.Chin}Chin S {2017 arXiv:1702.03219 [quan-ph]}
\bibitem{A.Streltsov}Streltsov A, Chitambar E, Rana S, Bera M N, Winter A and Lewenxtein M {2016 \emph{Phys.Rev.Lett.} \textbf{116} 240405}
%%%%????11-20
\bibitem{Y.Yao}Jedrzej Yao Y, Xiao X, Ge L and San C P {2015 \emph{ Phys. Rev. A } \textbf{92} 022112}
\bibitem{Y.Guo}Guo Y and Goswami S {2016 arXiv:1611.00413 [quan-ph]}
\bibitem{Chiranjib}Mukhopadhyay C, Das S, Bhattacharya S, Sen A and Sen U {2017 arXiv:1705.04343v1 [quan-ph]}
\bibitem{Y.Aharonov}Aharonov Y, Albert D Z and vaidman L {1988 \emph{Phys.Rev.Lett.} \textbf{60}(14)1351}
\bibitem{M.Koashi}Koashi M and Ueda M {1999 \emph{Phys. Rev. Lett.}\textbf{82}(12) 2598}
\bibitem{Haozhen}Situ H Z and Hu X Y {2016 \emph{Quantum Inf. Process}\textbf{15} 4649-4661}
\bibitem{ZouHM2}Zou H M and Fang M F {2015 \emph{Quantum Inf. Process} \textbf{14} 2673-2686}
\bibitem{Jaynes}Jaynes E T and Cummings F W {1963 \emph{Proc. IEEE} \textbf{51} 89}
\bibitem{ZouHM}Zou H M, Fang M F and Yang B Y {2013 \emph{Chin. Phys. B} \textbf{22} 12:120303}
\bibitem{zhm}Zou H M and Fang M F {2015 \emph{Chin. Phys. B \textbf{24} 8:080304}}
\bibitem{zhm2}Zou H M and Fang M F {2016 \emph{Chin. Phys. B \textbf{25} 7:070305}}
\bibitem{Bellomo1}Zou H M and Fang M F {2016 \emph{J. Mod. Optic} \textbf{63} 21:2279-2284}
%%%%????21-30
\bibitem{Bellomo2}Zou H M and Fang M F {2016 \emph{Chin. Phys. B} \textbf{25} 9:090302}
\bibitem{Breuer1}Breuer H P and Petruccione F 2002 \textit{The Theory of Open Quantum Systems}(Oxford: Oxford University Press) pp.~461-479
\bibitem{Ming}Wang M M and Qu Z G {2013 arXiv:1612.06020v1 [quan-ph]}
\bibitem{Scala1}Scala M, Militello B, Messina A, Maniscalco S, Piilo J and Suominen K A {2008 \emph{Phys. Rev. A} \textbf{77} 043827}
\bibitem{H.P.}Breuer H P, Laine E M and Piilo J {2009 \emph{Phys. Rev. Lett} \textbf{105} 050403 }
\bibitem{M.A.}Nielsen M A and Chuang I L 2000 \textit{Quantum Compution an Quantum Infor mation}(England: Cambridge University Press) pp.~399-400

\end{thebibliography}
\end{document}